\documentclass[aps, nofootinbib,reprint]{revtex4-1}

\usepackage{verbatim}

\def\beq{\begin{equation}}
\def\eeq{\end{equation}}
\def\ba{\beq\begin{array}{l}}
\def\ea{\end{array}\eeq}
\def\be{\ba}
\def\ee{\ea}

\def\p{\partial}

\def\u1{$U(1)_L$}
\def\ch{{\cal  H}}
\def\cp{{\mathcal P}}
\def\cq{{\cal  Q}}

\begin{document}

\title{Chiral Scale and Conformal Invariance in 2D Quantum Field Theory}

\author{Diego M. Hofman* and Andrew Strominger}
\email[Email:]{dhofman@physics.harvard.edu, andy@physics.harvard.edu}
\vskip.2in
\affiliation{Center for the Fundamental Laws of Nature, Harvard University, \\ Cambridge, MA 02138, USA }

%\date{\today}

\begin{abstract}
It is well known that a local, unitary Poincare-invariant 2D QFT with a global scaling symmetry and a discrete non-negative spectrum of scaling dimensions necessarily has both a left and a right local conformal symmetry. In this paper we consider a chiral situation beginning with only a left global scaling symmetry and do not assume Lorentz invariance. We find that a left conformal symmetry is still implied, while right translations are  enhanced either to a right conformal symmetry or a left U(1) Kac-Moody symmetry.

\end{abstract}

\maketitle

\section{Introduction}

  A two-dimensional (2D) Poincare and scale invariant quantum field theory (QFT) has at least four global symmetries
  under which the light cone coordinates transform as 
\be x^-\to x^-+a,~~~~~~x^+\to x^++b, \\ x^-\to \lambda^- x^-, ~~~~~~~~x^+\to \lambda^+ x^+.\ee
If in addition one posits that the theory is unitary and that the spectrum of the dilation operator for  $\lambda^+=\lambda^-$ is discrete and non-negative, then it was shown in \cite{joe} that the four global symmetries are enhanced to left and right infinite-dimensional conformal symmetries. Explicit counterexamples indicate that the enhancement need not occur if the dilation spectrum is not discrete. 

In recent years, interest in 2D QFTs with other types of global  scaling symmetries has arisen in a variety of contexts ranging from condensed matter to string theory.  In this paper we consider the special  case with three global symmetries:
 \be \label{ghj}x^-\to x^-+a, \quad x^+\to x^++b, \quad  x^-\to \lambda x^-
,\ee
comprising two translational symmetries and a chiral  ``left" dilational symmetry. Our assumptions, detailed in the next section, include locality, unitary and a discrete  non-negative  dilational spectrum but not Lorentz invariance. In an argument parallel to the one in \cite{joe}, we find that these three global symmetries are sufficient to conclude that there are (at least) two infinite-dimensional sets of local symmetries. One of these is a left local conformal symmetry which enhances the left dilational symmetry. The other enhances the right translational symmetry and can be either a right conformal symmetry or a left current algebra.

It is surprising that such a powerful conclusion can be reached from such minimal assumptions. However the element of surprise is potentially reduced by the fact that at this point there is no definite  example of a QFT which non-trivially satisfies all of our stated assumptions\footnote{One possible example might be given by the continuum limit of the large $N$ chiral Potts model, as discussed in \cite{Cardy:1992tq}.}.  Hence our powerful conclusions may apply to an empty set. On the other hand, possible interesting examples are suggested by the recent appearance of warped AdS$_3$ geometries in a variety of string theoretic investigations including the holographic duals of the so-called dipole deformations of 2D gauge theories \cite{dda,ddb,ddc,dde,ddd,ddf} and the near-horizon geometries of extreme Kerr black holes \cite{Bardeen:1999px,ghss,mgas}, see also \cite{Dijkgraaf:1996iy, Duff:1998us,Anninos:2008qb,Detournay:2010rh,Costa:2010cn}. Warped AdS$_3$ has an $SL(2,R)\times U(1)$  isometry group  which contains (\ref{ghj}). As these spaces are continuous deformations of AdS$_3$ spaces with CFT$_2$ duals, we expect that  their holographic duals exist and  are deformed CFT$_2$s with the symmetries (\ref{ghj}). However at present it not clear to what extent these duals obey all the assumptions stated below. In this paper we concentrate on the pure field theory analysis and leave these interesting issues to future investigations. It is also of interest to try to adapt our analysis to other types of 2D scaling symmetries such as, for example, Lifshitz scaling $x\to \lambda x$, $t\to \lambda^z t$. 

\section{From dilations to Virasoro}

We wish to consider local, unitary, translationally-invariant quantum field theories in 2D flat Minkowski space with a (linearly-realized) chiral global scale invariance. The assumed symmetries act on light-cone coordinates $x^\pm=t\pm x$ as 
\be x^-\to x^-+a,~~~~~~x^+\to x^++b, ~~~~~~x^-\to \lambda x^-, \ee
for constant $\lambda>0,a,b$.
We do not assume Lorentz invariance, an action or a conserved symmetric stress tensor. 
The operators generating left-moving (i.e. $x^-$) translations and dilations will be denoted $H $ and $D$ , while right-moving translations are denoted $\bar P $, where the bar in general denotes right-moving charges.  By assumption these operators annihilate the vacuum. 
Their commutation relations are:
\be\label{gol}
i \left[D, H\right] = H ,\quad i \left[D, \bar P\right] = 0,\quad i \left[H, \bar P\right] = 0. \, 
\ee
We moreover assume, following \cite{joe}, that the eigenvalue spectrum $\lambda_i$ of $D$ is discrete and non-negative\footnote{ For sigma-models this means the target space must be compact. } and there exists a complete basis of local operators $\Phi_i$ such that 
\be\label{discrete}
 i \left[H, \Phi_i\right] = \p_-\Phi_i, \quad i \left[\bar P, \Phi_i\right] = \p_+\Phi_i, \\
 i \left[D, \Phi_i \right] = x^-\p_-\Phi_i+\lambda_i \Phi_i 
\ee
and $\int_Cd\Phi_i=0$ for any closed or complete space-like contour $C$.  Note that  ``local operators" as here defined do not involve explicit functions of $x^\pm$. We will refer to $\lambda_i$ as the weight of the operator $\Phi_i$.
 Translational plus dilational invariance implies the vacuum two-point functions  of the $\{\Phi_i\}$  obey
\be\label{wt} \langle \Phi_i(x^-,x^+)\Phi_j(x'^-,x'^+)\rangle={f_{ij}(x^+-x'^+) \over (x^--x'^-)^{\lambda_i+\lambda_j}} \ee
for some a priori unknown functions $f_{ij}$. 

Noether's theorem implies  that each of the operators $H,~D,~\bar P$ is associated to a conserved Noether current with components denoted 
$h_\pm, d_\pm $ and $p_\pm$ whose dual contour integral, e.g. 
\be H=\int dx^+h_+-\int dx^-h_- \ee
then  gives the global charges.  A proof in the present context  is reviewed in appendix \ref{noether}.  All of these currents have an ambiguity under shifts of the form $\pm \p_\pm O$, where $O$ is a more general type of operator potentially involving  explicit functions of $x^\pm$
\be O(x^+,x^-)=\sum_i f_i(x^+,x^-)\Phi_i(x^+,x^-).\ee  We also show in appendix \ref{noether} that the shifts can be chosen so that  the currents satisfy canonical commutation relations, viz:
\be \label{hcurr} i[H,h_\pm]=\partial_-h_\pm,  \quad i[H,p_\pm]=\partial_-p_\pm, \\i[H,d_\pm]=\partial_-d_\pm  - h_\pm, \ee
\be \label{pcurr} i[\bar P,h_\pm]=\partial_+h_\pm, \quad  i[\bar P,p_\pm]=\partial_+p_\pm, \\ i[\bar P,d_\pm]=\partial_+d_\pm.  \ee
This implies that $h_\pm$ and $p_\pm$ are  local operators, but the term proportional to $h_\pm$ in $i[H,d_\pm]$ implies that $d_\pm$ must have explicit dependence on the $x^-$ coordinate.
Appendix \ref{noether} demonstrates - relying crucially on the discreteness of the spectrum $\lambda_i$ -  that the currents can be chosen to be eigenoperators of $D$.
The weights of the global charges (\ref{gol}) then imply  \be \label{fga} i[D, h_-]=x^-\p_- h_-+ 2h_-, \\ i[D, h_+]=x^-\p_- h_++h_+, \ee
 \be \label{fgb} i[D, p_-]=x^-\p_- p_-+ p_-, \\ i[D, p_+]=x^-\p_- p_+, \ee
\be \label{fg} i[D, d_-]=x^-\p_- d_-+ d_-,~~i[D, d_+]=x^-\p_- d_+. \ee We see that $d_+$ and $p_+$ are weight 0, $d_-$,  $h_+$ and $p_-$ are weight 1 and $h_-$ is weight 2.\footnote{In an ordinary  2D CFT , $d_+$ and $h_+$ and  $p_-$ all vanish,  $p_+=T_{++}$, $d_-=x^-T_{--}$ and $h_-=T_{--}$.}

As we mentioned, $d_\pm$ must have explicit coordinate dependence. Let us find it  and write the current in terms of local operators. Defining $s_\pm$ by 
\be\label{dilat1}
d_\pm = x^- h_\pm + s_\pm
\ee
and using (\ref{hcurr})  we find that \be \label{mc}
 i \left[H, s_\pm \right] =  \p_- s_\pm, \quad i \left[\bar P, s_\pm \right] =  \p_+s_\pm.
\ee
We conclude $(s_+,s_-)$ are local operators with weights  $(0,1)$. Conservation of the dilation current yields
\begin{eqnarray} \label{dilcon}
& &\partial_+ d_- + \partial_- d_+ =\\
  & &\quad x^- \left( \partial_+ h_-  +  \partial_- h_+\right) + h_+ + \partial_- s_+ + \partial_+ s_- = 0.\nonumber
\end{eqnarray}
 $h_\pm$ conservation then gives
\be \label{dilcon2}
h_+ = -  \partial_- s_+ - \partial_+ s_-
\ee

We have not at this point fixed all the shift freedom in the currents.  In particular, we may shift away  $s_-$:
\be\label{repcur}
h_\pm \rightarrow h_\pm \mp \partial_\pm s_-, \quad d_\pm \rightarrow d_\pm \mp \partial_\pm \left(x^- s_-\right), 
\ee
which remains consistent with the commutators (\ref{hcurr}-\ref{fg}) as well as current conservation. 
Equations (\ref{dilat1}) and (\ref{dilcon2}) now take the simpler form
\be \label{dilat2}
 d_+ = x^- h_+ + s_+, \quad\quad d_- = x^- h_- 
\ee
and
\be\label{dilcon3}
h_+ = - \partial_- s_+
\ee

Now, we can use the general form of the two-point functions given by (\ref{wt}) . Bearing in mind the fact that $s_+$ is a local operator of  weight 0,  we must have \be\label{O2}
\langle s_+ s_+ \rangle = f_{s_+}(x^+)
\ee
which implies \be \partial_- s_+ =h_+=0. \ee Conservation of  $h_\pm$ then reduces to $\partial_+ h_-=0$ or, equivalently, 
\be\label{hpm}
 h_-=h_-(x^-).
\ee
This fact immediately leads to  the existence of an infinite set of conserved charges. 
Define:
\be\label{vira1}
T_\xi = -\int dx^- \, \xi(x^-)  h_-, \quad \ \bar{J}_\chi = \int dx^+ \, \chi(x^+)  s_+
\ee
 where $\xi(x^-)$ and $\chi(x^+)$ are smooth functions. In particular, $H=T_1$ and $D=\bar{J}_1 + T_{x^-}$.  Notice that, while $h_-$ can't vanish if we are to have a non trivial $H$ operator, $s_+$ could be identically zero.  $s_+\neq 0$ leads to the existence of even more local symmetries unrelated to the originally posited global symmetries. Whether or not it is zero, we show in appendix \ref{splus} that 
\be
i [\bar{J}_\chi, T_\xi] = 0.
\ee
This means we are free to calculate the algebra spanned by the conserved charges $T_\xi$ without worrying about the action of $\bar{J}_\chi$.  We henceforth concentrate on the minimal case  $s_+=0$.

Let us now work out the algebra spanned by $T_\xi$. The action of $H=T_1$ and $D=T_{x^-}$ on $h_-$ imply
\be
i[T_1, T_\xi] = T_{-\xi'} ,\,\, i[T_{x^-}, T_\xi] = T_{\xi-\xi' x^-},  \,\,  \xi'\equiv\partial_- \xi.
\ee
This in turn  implies that the action of $T_\xi$ on $h_-$ is 
\be\label{tono}
i[T_\xi, h_-] = \xi \partial_- h_- + 2 \xi' h_- + \partial_-^2 O_\xi.
\ee
The scaling symmetry plus locality implies $O_\xi$ must be of the form $O_\xi=\xi O_1+\xi'O_0$ with $O_1$ a local operator of weight one. As it cannot depend on $x^+$,  $O_0$ must be a weight zero constant. Integrating both sides with respect to  $dx^-\zeta(x^-)$ gives 
\be
i [T_\xi, T_\zeta] = T_{\xi' \zeta-\zeta' \xi} + \int dx^- \,  \zeta\p_-^2O_\xi.
\ee
Antisymmetry under the exchange of $\xi \leftrightarrow \zeta$ then requires that $O_1=0$ and :
\be
\partial_-^2 O_\xi  = O_0 \partial_-^3 \xi.
\ee
Defining $c=24\pi O_0$ we end up with the following commutations relations for the charges $T_\xi$:
\be
i [T_\xi, T_\zeta] = T_{\xi' \zeta-\zeta' \xi} + \frac{c}{48\pi} \int dx^- \, (\xi'' \zeta' - \zeta'' \xi').
\ee
We recognize this as the algebra of the left-moving conformal generators on the Minkowski plane with central charge $c$.

\section{Enhancement of right moving translations to a local symmetry}

Notice that up until now we have not made much use of the fact that our theory is translationally invariant in the $x^+$ dimension as well. In particular the above results also apply when we do not posses this symmetry. What happens now that we add $p_\pm$ to the game?

In this case, the key observation is that $p_+$ is a zero weight local operator, as implied by (\ref{hcurr}-\ref{fg}). This means
\be\label{pm2}
\langle p_+ p_+ \rangle = f_{p_+}(x^+).
\ee
Acting with  $\p_-$ on both insertions  and using the fact that a hermitian operator with a vanishing  two-point function is trivial we learn that $\partial_- p_+=0$. Current conservation then implies  $\partial_+ p_- =0$. It follows that 
\be\label{ppm}
p_+ =p_+(x^+) ,\quad\quad p_-=p_-(x^-).
\ee
We cannot have both $p_+=0=p_-$ as the charge $\bar P$ is generically nonzero, although 
from what we have seen so far one of them could vanish. 
We now discuss  all possibilities.
\subsection{$p_-=0 \Rightarrow $ right-moving Virasoro algebra}In this case we have infinitely many charges 
given by
\be\label{vira2}
\bar{T}_\xi = \int dx^+ \, \xi(x^+)  p_+.
\ee
Since $\bar{T}_1=\bar P$ we have
\be
i[\bar{T}_1, \bar{T}_\xi] = -\bar T_{\xi'}.
\ee
This, in turn, constraints the action of $\bar{T}_\xi$ on $p_+$ to be 
\be\label{tono2}
i[\bar{T}_\xi, p_+] =\xi \partial_+ p_+ + 2 \xi' p_+ + \partial_+ \bar{O}_\xi.
\ee
 If we compare this expression with (\ref{tono}) we see that we are very close to the previous situation for $T_\xi$. Multiplying by $\zeta(x^+)$  and integrating both sides of this equation we get 
\be\label{too2}
i[\bar{T}_\xi,\bar{T}_{\zeta}   ] =\bar{T}_{\xi \zeta'-\xi'\zeta} + \int dx^+ \zeta \partial_+ \bar{O}_\xi.
\ee
Antisymmetry with respect to exchange of $\xi$ and $\zeta$ then implies that $\bar{O}_\xi$ is an even number of derivatives of $\xi$. The term with no derivatives can be eliminated by a constant shift of $p_+$.  Terms with four or more derivatives would violate the Jacobi
identity.  We conclude (shifting $p_+$ by a constant if needed)
\be
i [\bar T_\xi, \bar T_\zeta] = \bar T_{\xi' \zeta-\zeta' \xi} + \frac{\bar c}{48\pi} \int dx^+ \, (\xi'' \zeta' - \zeta'' \xi').
\ee
We recognize this as the algebra of the right-moving conformal generators on the Minkowski plane with central charge $\bar c$. \footnote{Interestingly, it is this right moving Virasoro that gives the entropy in Kerr-CFT\cite{ghss}.  }

Of course the vacuum will not in general be invariant under the global $SL(2,R)_R$ subgroup. 
Acting  with $D$ and $H$ on $\bar{T}_\xi$ we can check that $i[\bar{T}_\xi, h_-]=\partial^2_- \Phi_\xi$. But $\Phi_\xi$ must be a weight zero operator, se we are left with $i[\bar{T}_\xi, h_-]=0$. The upshot  is that $[\bar{T}_\xi, T_\zeta]=0$, as expected.

\subsection{ $p_+=0=>$ left-moving current algebra} 
In this case we have infinitely many left-moving charges 
\be 
\quad \quad J_\chi =- \int dx^- \, \chi(x^-)  p_-(x^-).\ee
Because the zero mode $J_1$ acts as $\partial_+$, we must have $i[J_1,p_-]=0$. This implies $[J_1,J_\chi]=0$ and hence
\be
i[J_\chi,p_-]=\partial_- M_\chi
\ee
where the operator $M_\chi$ is, by locality, a linear function of $\chi$. Now we are in position to repeat the arguments used around (\ref{tono}).  Multiplying by $\psi(x^-)$,  integrating over $x^-$ and invoking antisymmetry and the Jacobi identity we find
\be
i[J_\chi, J_\psi] = \frac{k}{4 \pi} \int dx^- (\psi' \chi - \chi'\psi)
\ee
where the constant $k$ parameterizes the central element. This is a $U(1)$ Kac-Moody current algebra.  

We also need the $[T_\xi,J_\chi]$ commutator. The fact that 
$[J_1,T_\xi]$=0 implies
\be [T_\xi,p_-]=\xi\p_-p_-+\xi'p_-+\p_-N_\xi \ee
with the operator $N_\xi$ linear in $\xi$. The Jacobi identity with a third operator $T_{\zeta}$  then implies $N_\xi=f \p_-\xi$ for some constant $f$  \cite{Polyakov:1988am,daCosta:1993hg}. If $f$ is nonzero, the current $p_-$ is not a dimension one chiral current. However we may then shift $h_-$ by a $k$-dependent multiple of  $\p_- p_-$ so that $p_-$ is a good dimension one current. This shift affects the central charge of $T_\xi$. Performing this transformation leaves us with the standard commutator
\be
i[T_\xi, J_\chi] = J_{-\xi \chi'}.
\ee

It may seem rather strange to have a left-moving Kac-Moody current algebra whose zero mode generates right translations. However reminiscent structures have appeared before.  In the KK circle  reduction of AdS$_3$ to AdS$_2$, one begins with two Virasoros and ends with a single left-moving Virasoro and current algebra  associated to the KK U(1)  \cite{ads2}. The left current algebra zero mode $J_0$ descends from right Virasoro zero mode in AdS$_3$. Related structures have appeared in Kerr/CFT, where left Virasoro and right current algebra zero modes are sometimes identified \cite{Bredberg:2009pv},  as well as in the study of asymptotic symmetries of warped AdS$_3$  \cite{cdt}. 
\subsection{Non-minimal $p_-\neq 0, ~~p_+ \neq 0$}

In this case left and right currents decouple. The commutators $i [ \bar P, \bar{T}_\xi] = \bar{T}_{-\xi'}$ implies $i [ \bar{T}_\xi , p_-] = \partial_- \Phi_\xi$, for some local $\Phi_\xi$. Furthermore, because $\bar{T_\xi}$ does not transform under $D$ and $p_-$ is a weight 1 operator,  $\Phi_\xi$ must be weight 0. Therefore
\be
 i[\bar{T}_\xi, J_\chi]=0.
\ee
implying that the conserved charges can be analyzed separately as above.

In summary,  left translational and dilational symmetries together with right translations imply the existence of (at least) two sets of infinite dimensional algebras. On the left we always find a local conformal symmetry, while the right translational current is enhanced either to a local right conformal symmetry or  a left U(1) current algebra.

\begin{acknowledgements}
We are delighted to acknowledge fruitful discussions with C. Bachas, A. Castro, M. Guica, T. Hartman, S. Hartnoll, J. Maldacena, A. Maloney, J. McGreevy, J. Polchinski, S. Sachdev, N. Seiberg and W. Song. This work was supported in part by DOE grant  DE-FG0291ER40654 and the Fundamental Laws Initiative at Harvard.
\end{acknowledgements}

\appendix
\section{ Noether's theorem}\label{noether}
In this section we will prove Noether's theorem for $H$ and $P$ and show the Noether currents can be put in a canonical ``diagonal" form.  We assume the existence of a unitary hamiltonian $\ch$ whose commutator with any operator obeys \be\label{com1}
i[\ch, O] = \frac{ d O}{d t}- \frac{\partial O}{\partial t} ,
\ee
where the last derivative acts on any explicit coordinate dependence in $O$.
Conserved charges are defined as any operator $\cq$ such that $\frac{ d \cq}{d t}=0$. Locality implies 
\be \cq =\int^\infty_{-\infty} dx \, q_t(x,t). \ee
Charge conservation is then 
\be
\frac{d \cq}{d t} = \int dx \, \frac{ q_t}{d t} = 0,
\ee
implying  $\frac{ q_t}{d t} = - \frac{ q_x}{d x}$, for some $q_x$. $(q_x,q_t)$ is the sought after  conserved Noether current associated to $\cq$. 

In this paper we further assume the existence of a conserved momentum charge $\cp$ commuting with  $\ch$  and obeying 
\be\label{com2}
i[\cp, O] = \frac{ d O}{d x}- \frac{\partial O}{\partial x}
\ee
for any operator $O$. From these we construct left and right translation charges $2H=\ch-\cp$
and $2\bar P= \ch+\cp$.  If a conserved charge $\cq$ commutes with both $H$ and $\bar P$ the associated Noether current must obey 
\begin{eqnarray}
i[H, q_\pm] & =& \partial_- q_\pm \pm \partial_\pm F\label{hq1},\\
i[\bar P, q_\pm] &=& \partial_+ q_\pm \pm \partial_\pm G\label{pq1},
\end{eqnarray}
\noindent where $\partial_\pm$ are total derivatives with respect to $x^\pm=t\pm x$ and $F$ and $G$ can be expanded \be
F = \sum_i f_i(x^+,x^-) \Phi_i, \quad G=\sum_i g_i (x^+,x^-) \Phi_i.
\ee
The Jacobi identity relates the coefficients $f_i$ and $g_i$
\be\label{jacobi}
[[H,\bar P],q_\pm] = [H,[\bar P,q_\pm]]-[\bar P,[H,q_\pm]] =0.
\ee  This translates into  an integrability condition implying the existence of a set of functions $r_i$ such that $g_i = \partial_+ r_i$ and $f_i=\partial_-r_i$.
Now let us use the  shift freedom $q_\pm  \rightarrow q_\pm \mp \partial_\pm R$, with $R = \sum_i r_i(x^+,x^-) \Phi_i$. We then obtain the canonical form of the commutators
\be
i[H,q_\pm]=\partial_- q_\pm, \quad\quad i[\bar P,q_\pm]=\partial_+ q_\pm.
\ee
We see  that any current associated to a symmetry generating charge that commutes with the hamiltonian and the momentum operator can be chosen so as to have no explicit dependence on the coordinates. In particular, this applies to the $h_\pm$ and $p_\pm$ current themselves. It is worth pointing out that, for any local (coordinate independent) operator $\Phi$ we can still shift $q_\pm \rightarrow q_\pm \mp \partial_\pm \Phi$ and preserve the above commutation relations.

Now we show, following \cite{joe}, that the currents can also be made dilation eigenoperators. Current conservation and the commutation relations (\ref{gol}) imply:
\be\label{Dhp1}
i[D, h_\pm] = x^-\partial_- h_\pm + \lambda \left(h_\pm\right) h_\pm \pm \partial_\pm O_h ,\\
 i[D, p_\pm] = x^-\partial_- p_\pm + \lambda \left(p_\pm\right) p_\pm \pm \partial_\pm O_p ,
\ee
where $\lambda \left(p_+,p_-,h_+,h_-\right)=\left(0,1,1,2\right)$. The Jacobi identity can be used to show  that $O_h$ and $O_p$ are local operators with no explicit coordinate dependence.  This means they are expandable in the (by assumption) discrete basis (\ref{discrete}) 
\be
O_q = \sum_i a_i \Phi_i
\ee
for $q=h,p$ where $\Phi_i$ has weight $\lambda_i$.   Let us now shift 
\be q_\pm \rightarrow q_\pm \mp \partial_\pm \sum_i b_i \Phi_i ,~~~~~b_i
= \frac{a_i}{w(q_+)-\lambda_i},\ee for  $\lambda_i \neq w(q_+)$. This shift eliminates all the  $\Phi_i$ in $O_q$ with weights $\lambda_i \neq w(q_+)$.  Operators with weight equal to $w(q_+)$ cannot appear in $O_q$ by the assumption (\ref{discrete}) that the spectrum of $D$ is discrete and diagonalizable. We are therefore left with the canonical form of the commutators:
\be\label{Dhp2}
i[D, h_\pm] = x^-\partial_- h_\pm + \lambda \left(h_\pm\right) h_\pm,\\
 i[D, p_\pm] = x^-\partial_- p_\pm + \lambda \left(p_\pm\right) p_\pm.
\ee

It was shown above that there  exists a conserved current $d_\pm$ associated to $D$.  A slight variant of the preceding arguments shows that we can exploit the shift freedom to set\be\label{dilatcan}
i[H, d_\pm] = \partial_- d_\pm - h_\pm, \quad\quad i[H, d_\pm] = \partial_+d_\pm.
\ee
The non trivial commutator $i[D,H]=H$ requires the extra term proportional to $h_\pm$, which implies that the dilation current cannot be independent of the coordinates.

Finally, the action of the dilation charge on its own associated current can be obtained as above and shifted to the canonical form  
\be\label{Dd}
i[D, d_\pm] = x^-\partial_- d_\pm + \lambda \left(d_\pm\right) d_\pm, 
\ee
where $\lambda \left(d_+,d_-\right)=(0,1)$.
We note that 
demanding that the current/charge commutators take this canonical form does not fix all the ambiguity  in the former. Some of the remaining  shift freedom is exploited in this work around equation (\ref{repcur}) to shift away $s_-$. \\

\section{Decoupling of $s_+$}\label{splus}

In this appendix we show that in the non-minimal case $s_+\neq 0$ , $[\bar{J}_\chi , T_\xi]=0$. Given that $s_+$ is independent of $x^+$ and transforms as a zero weight operator under $D$, we have
\be\label{dilaschi}
\begin{array}{c}
i[H, s_+] =0 \\
 i[D,s_+]=0
\end{array}
  \rightarrow 
  \begin{array}{c}
  i[H, \bar{J}_\chi] =0 \\
  \,i[D,\bar{J}_\chi]=0.
     \end{array}
\ee
The Jacobi identity implies that the commutator of an operator annihilated by $H$ (such as $\bar{J}_\chi$) and a local field (such as $h_-$) must be a local field itself. Therefore, using (\ref{dilaschi}), we have 
\be
i[\bar{J}_\chi, h_-] = \partial_-^2 \Phi_\chi,
\ee
where $\Phi_\chi$ is a local operator of weight zero. This implies, $\partial_- \Phi_\chi=0$ as an operator equation. Immediately we get
\be
i [\bar{J}_\chi, T_\xi] = 0.
\ee

\end{document}